\documentclass{article}
\usepackage{graphics}
\usepackage{graphicx}
\usepackage{epstopdf}
\usepackage{subfig}
\begin{document}
\title{The `Sears paradox'}
\author{Jerrold Franklin\footnote{Internet address:
Jerry.F@TEMPLE.EDU}\\
Department of Physics\\
Temple University, Philadelphia, PA 19122-6082}
\maketitle
\begin{abstract}
We resolve a paradox in special relativity proposed by F. W. Sears for the action of forces on a rigid body.
In the paradox, a moving rigid rod is struck at different times by impulsive forces, but continues to move with unchanged velocity, with the forces having no effect.
Our conclusion is that the usual laws of mechanics can be applied to a rigid body only in its rest system. 

\end{abstract}

\section{The Sears paradox}

A paradox concerning rigid body interactions was described by F. W. Sears some time ago\cite{sears}.
He considered a rigid rod at rest in a Lorentz frame S.  The rod is simultaneously acted on at each end by collinear equal and opposite time dependent forces.
Since the net force is zero, the rod will remain at rest in frame S.  
In a frame S$'$, moving with constant velocity $-v$ with respect to S,  the rod will have a constant velocity $+v$.  However, due to the relativity of simultaneity, the two forces at the ends of the rod will now be equal at different times.  The paradox is, {\em How can the velocity of the rod remain constant in frame S$'$  where the forces are no longer equal and opposite at the same time?}  
Although this paradox was posed by Sears many years ago, somewhat surprisingly there do not seem to be any subsequent papers addressing the issue he raised.
In this paper,  we resolve the question of what really happens in the frame of the moving rod.

We consider a simplified example where the forces are due to equal and opposite impulses caused by elastic collisions of point masses with each end of the rod,
with all motion and the orientation of the rod along the x-axis.
This situation is illustrated in space-time diagrams in Fig.~1.
In Fig.~1a,  a rod of length $L$ and mass $M$ is originally at rest in a Lorentz frame S.
For simplicity, the point particles each have the same mass $M$ as the rod, and initial speed $v=0.6$ (We use units with $c=1$.) directed toward either end of the rod, 
particle A from the left, and particle B from the right.
At time $t=0$ in frame S, each point particle is a distance $L/2$ from an end of the rod of length $L$. 

Figure 1a shows the path of each point object before and after collision with the rod, which remains stationary because the simultaneous equal and opposite impulses cancel.
The equations for this and the other figures are given in the Appendix.
\begin{figure*}[h]
\centering
\begin{minipage}[b]{0.45\linewidth}
\includegraphics[height=3.5in]{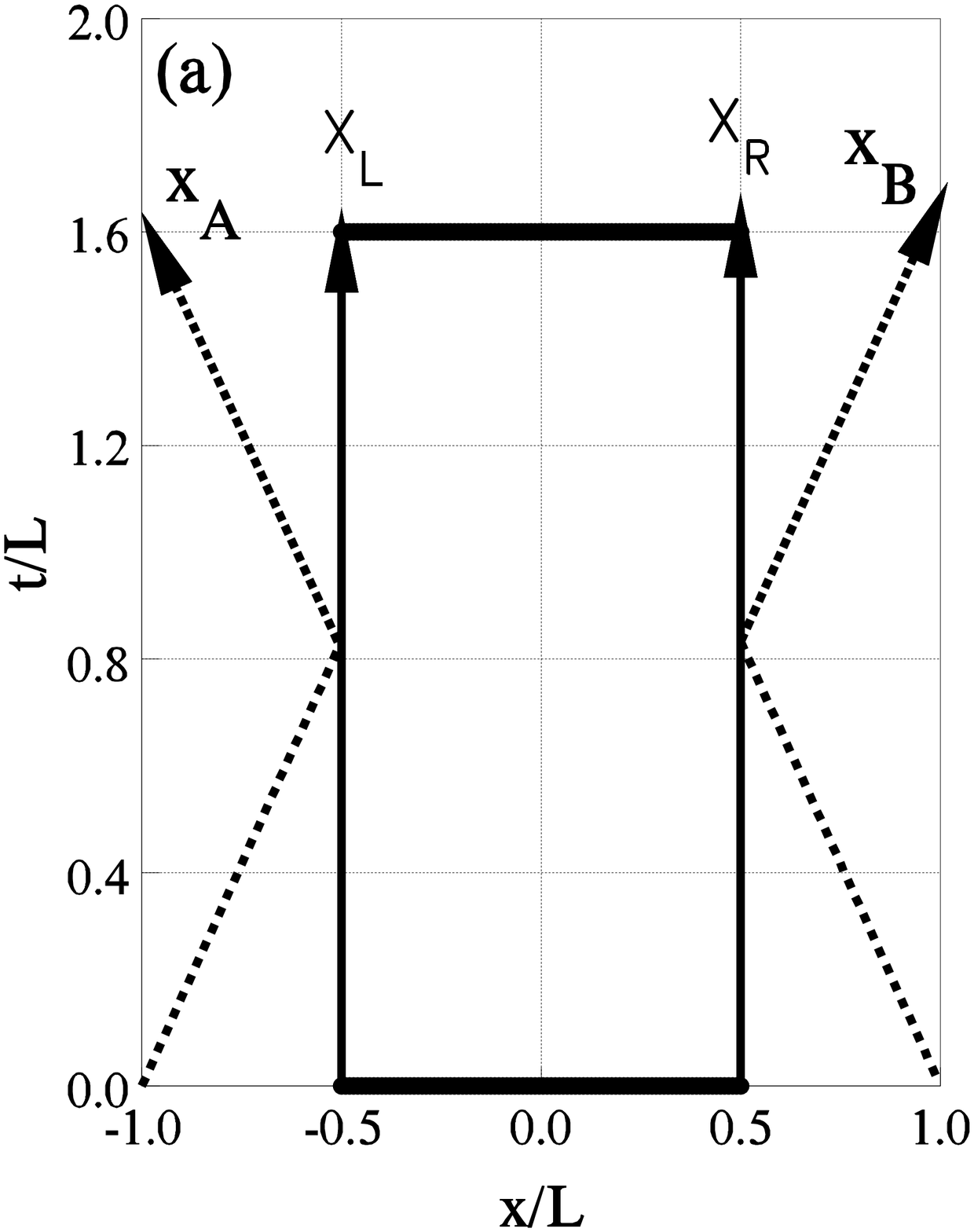}
\end{minipage}
\quad\quad
\begin{minipage}[b]{0.45\linewidth}
\includegraphics[height=3.4in]{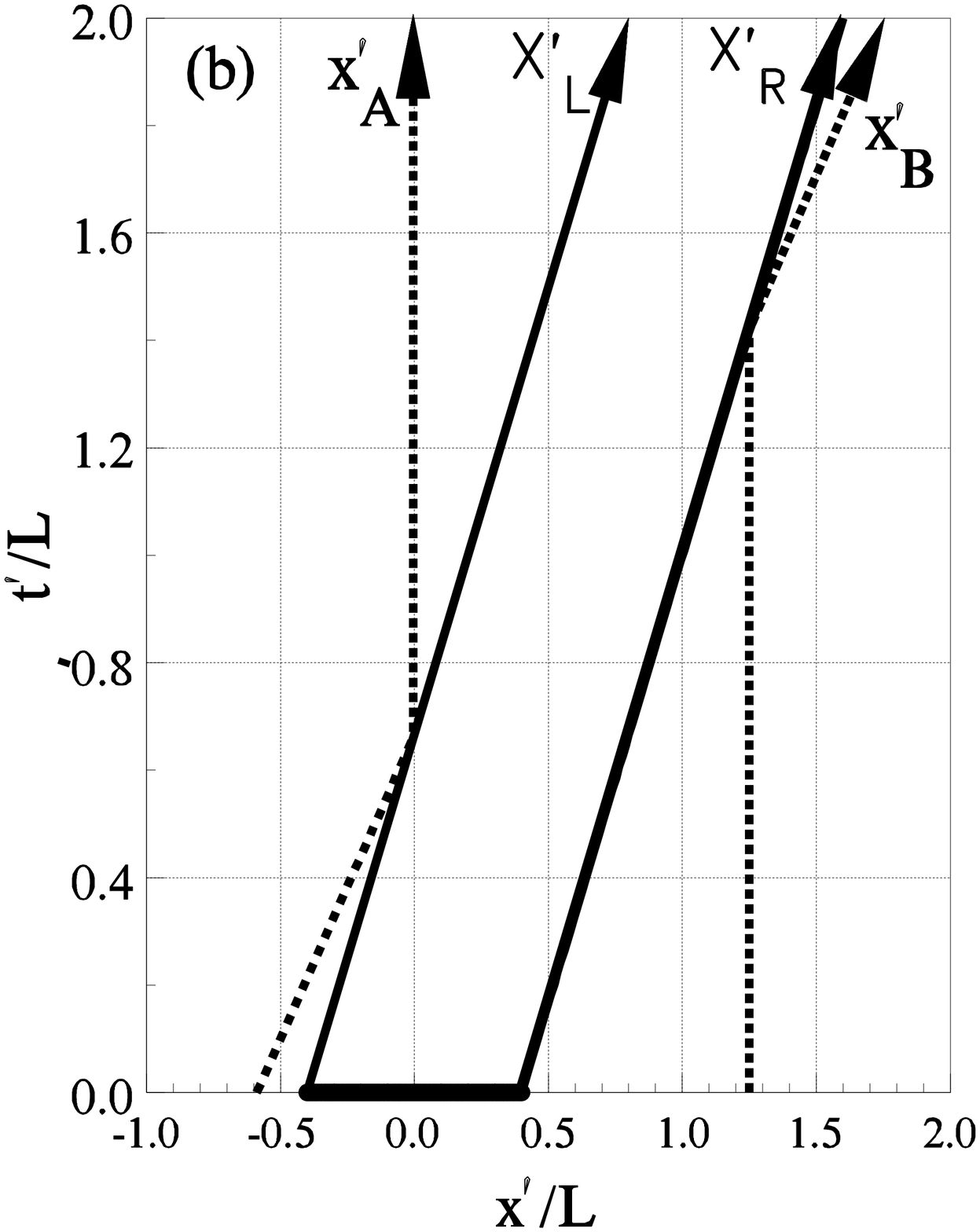}
\end{minipage}
\caption{Impulsive collisons by particles A and B (dashed lines) on the ends of a rigid rod (solid lines).
(a) Simultaneous impacts by particles A and B in the rest system S of the rod.
(b) Trajectories in system S$'$ where the rod has velocity $v$.  The impacts of particles A and B are no longer simultaneous, but the rod continues in a straght line with constant velocity.}
\end{figure*}

The trajectories in the rod's rest frame can be Lorentz transformed to a frame S$'$ that moves with velocity $-v$ with respect to frame S.
The trajectories in frame S$'$ are shown in Fig.~1b. 
As the figure shows, in frame S$'$ particle A is the first to strike the rod.
However, the motion of the moving rod does not change because a constant velocity path in one Lorentz frame Lorentz transforms into a constant velocity path in any other Lorentz frame.  Once the trajectories are found in the rest frame of the rod,  they are uniquely determined in any other Lorentz frame.  The paradox is that,
if momentum were conserved in the first collision in system S$'$, the rod would be bumped forward until being hit again by particle B.  But this doesn't happen.
 
Sears attempts to resolve this paradox with the concept of  `hidden momentum', by which the rod gains momentum that does not affect its velocity.  When the rod is later struck by particle B, this hidden momentum is imparted to particle B, and again the velocity of the rod is unchanged.
There are several objections to this resolution.  How does the rod in S$'$ know it is going be hit at some future time?  This would require two kinds of elastic collisions with the rod.  One, with the normal relation between momentum and velocity, and another collision with the same initial impact, but with a hidden connection between momentum and velocity.  How does one decide which type of collision takes place?  Also, what if the second object missed the rod.  Would the rod have to have known this and recoiled at the first collision? 

The key to resolving the paradox is to realize that there can only be one Lorentz frame in which the usual condition for equilibrium of forces can be valid.
This is because balancing forces that act simultaneously in one Lorentz frame will not be simultaneous in another.  
Yet, if a trajectory has constant velocity in one Lorentz frame,  it must have constant velocity in any other Lorentz frame.
This means there can be only one Lorentz system in which the usual dynamical force equations apply.
It seems clear to us that this system must be the rest system of the rod.
That is, the forces that move an object (or keep it in equilibrium) act on the object in its rest system.  
To get anthropomorphic about it, when you feel a push you are in your rest system.
The dynamics of ${\bf F}=\frac{\bf dp}{dt}$ acts on you only in your rest system, and determines your trajectory in that system.  
Your trajectory in any other Lorentz system is then determined by a Lorentz transformation of the trajectory to that system.   

In the example Sears proposed, this means that the moving rod continues at constant velocity in system S$'$ even though it appears to be struck by the opposing forces at different times .  Sears agrees with this, but we reject his proposal that `hidden momentum' enters the rod without changing its velocity.  Our proposal is that the time sequence seen in the system where the rod is moving is an illusion caused by the Lorentz transformation.  The illusion results from the fact that, in relativity, events that are simultaneous in one Lorentz system will not be simultaneous in another.   The two forces that appear to occur at different times in frame S$'$ are actually simultaneous in the rod's rest frame.

This effect can be seen as a rotation in 4D space-time
that is similar to a non-relativistic 3D rotation.  If a rod at rest is rotated about a vertical axis, two forces exerted at the same time on each end will appear to be acting at differing times. This is because there are different time delays in the passage of light from each end.  Thus, the illusory effect (apparent lack of simultaneity) caused by a rotation in 4D space-time is similar to that caused by a rotation in 3D space.  This explains why the action of the forces on each end of the moving rod appear to be at different times, even though they are simultaneous in the rest system.  We note that if the particles struck the ends of the moving rod at any other times than the specific times in Fig.~1, the rod would recoil.  We discuss an example of this in the next section. 

\section{A related case}

An interesting situation arises if we try to act on the moving rod in system S$'$ with equal and opposite simultaneous impulses.  
We illustrate this in Fig.~2.  In Fig.~2b, particle A is now initially positioned somewhat to the left of where it was in Fig.~1b.  In this position, it strikes the rod at the same time that particle B strikes the rod from the right. 
The initial positioning of particle A in Fig. 2b is somewhat complicated by the fact that the left end of the rod starts to slow down before its right end strikes particle B. The trajectory of the left end of a rod when the right end of the rod is brought to rest by an impulsive impact
is described in section 5 of Ref. 2, and the equations for this are given in the Appendix of this paper.  We note that, because the left end of the rod starts to slow down before the right end is stopped by striking particle B, the rod attains its rest length when it comes to a momentary halt, as shown in Fig. 2b.

\begin{figure*}[h]
\centering
\begin{minipage}[b]{0.45\linewidth}
\includegraphics[height=3.3in]{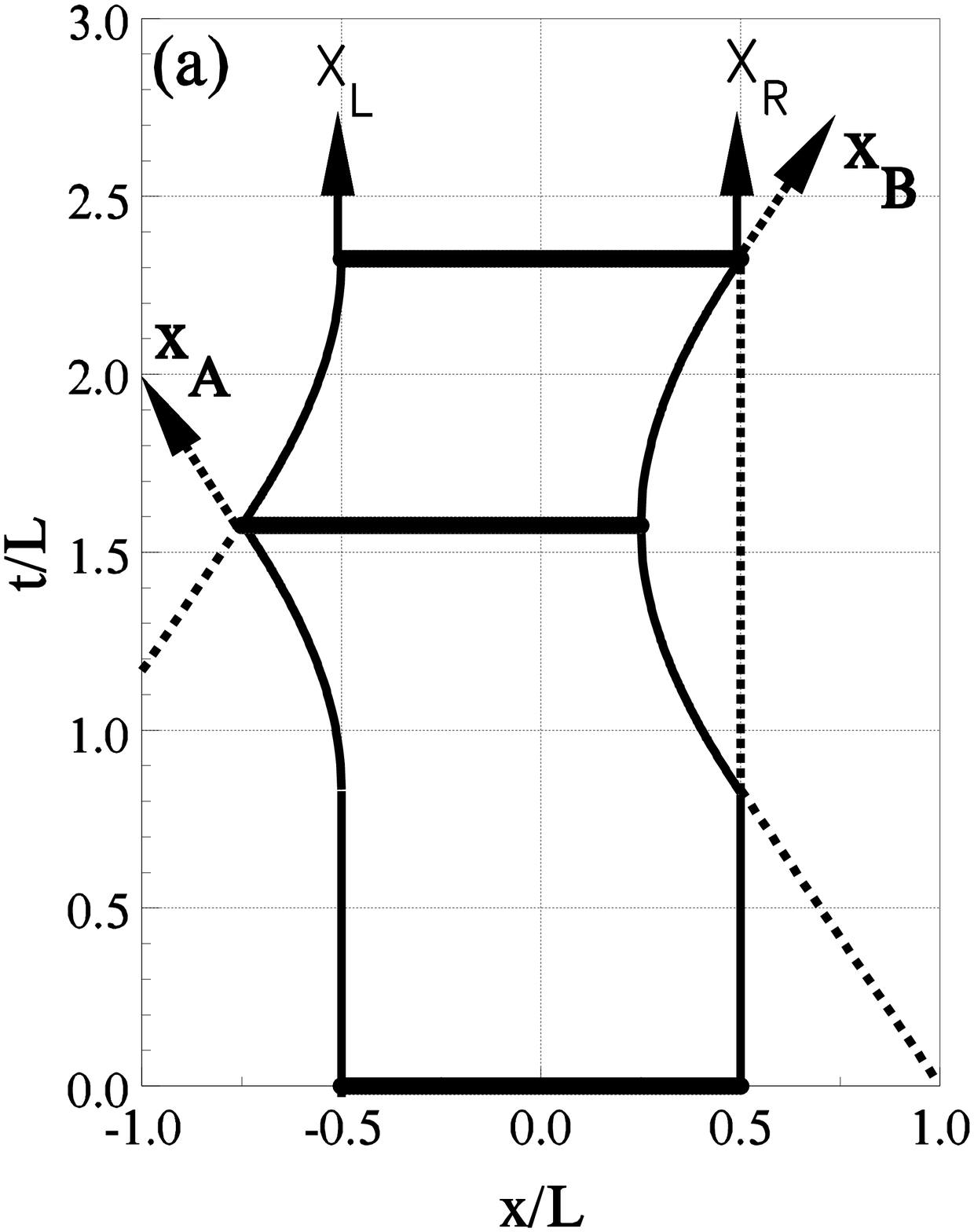}
\end{minipage}
\quad\quad\quad
\begin{minipage}[b]{0.45\linewidth}
\includegraphics[height=3.4in]{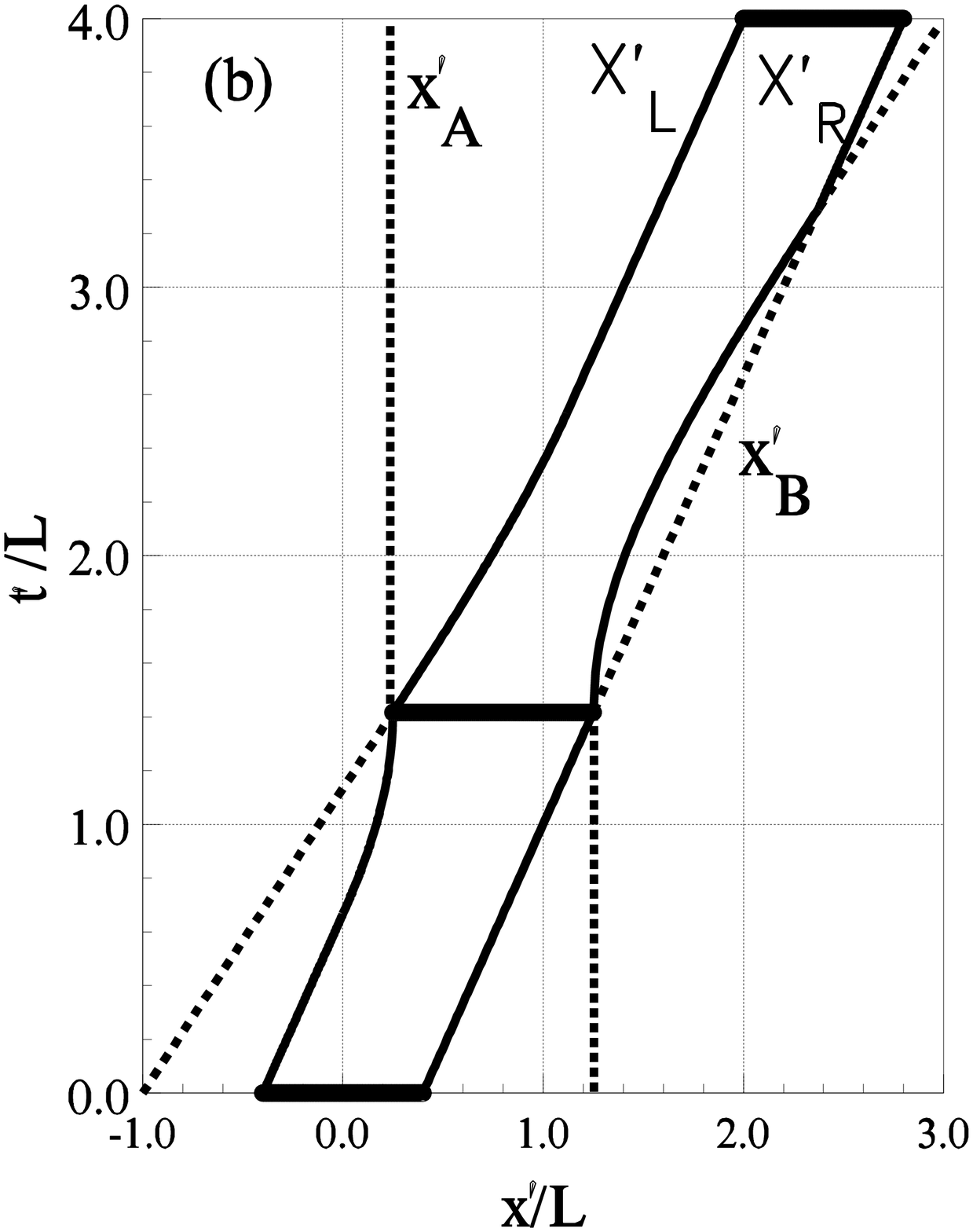}
\end{minipage}
\caption{Impulsive collisons by particles A and B (dashed lines) on the ends of a rigid rod (solid lines).
(a) The impacts of particles A and B are not simultaneous in the rest system S of the rod, 
and the rod follows the zig-zag trajectory shown.
(b) The first impacts by particles A and B are simultaneous in system S$'$ where the rod has velocity $v$.  However
 the rod does not continue with constant velocity, but follows the more complicated trajectory shown.}
\end{figure*}
     
In system S, where the rod is initially at rest, 
 particle B strikes the rod before particle A
reaches the left end of the rod.  
The rod will thus follow the zig-zag trajectory shown in Fig.~2a.  It first reacts to the right hand impulse, then rebounds when it is struck by particle A, 
and finally comes to rest again when it is struck a second time by particle B.  The equations for the trajectory followed in system S by the rigid rod before coming to rest again after the three impacts have been derived in Ref.~\cite{jfr}, and are given in the Appendix.

The trajectories in system S$'$ are also unusual.   While these trajectories are completely determined by Lorentz transformation from system S, their appearance in system S$'$ can be described as follows.  The rod is initially moving to the right with speed $v$, and has Lorentz contracted length $L/\gamma $. Then (as described in Ref.\cite{jfr}), the left end of the rod starts to slow down before the right end of the rod strikes the right hand particle B, which is initially at rest.  The entire rod comes to rest for a moment at the time the right end strikes particle B.  At that point the rod has its rest frame length $L $, which was achieved because the left end started to slow down before the right end stopped.  
 
Then particle A impacts the left end of the rod with velocity $v'=2v/(1+v^2)$.
We see in Fig.~2b that the initial impact of the two particles on the moving rod are simultaneous, but these impulses do not cancel  because they are not simultaneous in the rod's rest system.
Following the impact by particle A, the left end of the rod
starts to move at speed $v'$, while the right end starts more slowly so that the rod eventually attains its contracted length when the entire rod reaches speed $v'$. The right end of the rod then hits the now stationary particle B, with the left end of the rod again starting to slow down first.  After this final collision, the rod continues with its original velocity, $v$, and the two particles continue with their original velocities interchanged.

\section{Summary}

We have seen that Newton's first law of motion can only be  implemented in the rest frame of a rigid body.
That is, a rigid body that is at rest will remain at rest if the simultaneous forces acting on it have a zero resultant.
If the rigid body is in motion with a constant velocity, it will maintain that velocity only if the forces acting on it have a zero resultant when they are simultaneous in the rest frame of the rigid body.  We have shown that a rigid body in motion with constant velocity that is acted on by 
equal and opposite impulses that are simultaneous in its moving frame does not maintain its constant velocity.

We have shown in a previous paper\cite{jfr} that Newton's  second law $({\bf F = dp/} dt)$
 can only be applied to a rigid body in its rest frame.  If the rigid body is in motion, then repeated Lorentz transformation to its instantaneous rest frame must be implemented in applying Newton's second law. We have only shown this for the case of acceleration that is constant in time. We believe that the general principle would hold for a time varying acceleration, but the trajectory equations would be more complicated.   In the present paper, we have applied the equations from Ref.\cite{jfr} to show that  
 a rigid body in motion acted on by 
equal and opposite impulses that are simultaneous in its moving frame follows the zig-zag trajectory of fig.~2.

Our conclusion is that the usual laws of mechanics can be applied to a rigid body only in its rest system.

\section{Appendix}

In this Appendix, we present the equations used for the figures in the paper.\\  
\\
{\bf Fig.~1a:}\\  
For Fig.~1a, the left end $X_L$ and the right end $X_R$ of the rod are at rest, so
\begin{equation}
X_L=-L/2,\quad X_R=+L/2.
\label{eq:rod}
\end{equation}
Particles A and B move at velocities plus and minus $v$ until they strike the rod at time $t=L/2v$, so
\begin{equation}
x_A=-L+vt,\quad x_B=+L-vt,\quad t<L/2v,
\label{xLR}
\end{equation}
where $x_A$ and $x_B$ are the positions of particles A and B.
After the impact at $t=L/2v$, their velocities are reversed, and
\begin{equation}
x_A=-vt,\quad x_B=+vt,\quad t>L/2v.
\end{equation}
\\
{\bf Fig.~1b:}\\
In Lorentz frame S$'$, the rod has velocity  $v$ and Lorentz contracted length $L/\gamma$, while particle A has initial velocity
\begin{equation}
v'=\frac{2v}{1+v^2},
\end{equation}
and particle B is initially at rest.

The trajectories of the two particles, $x'_A(t')$ and $x'_B(t')$, and $X'_L(t')$ and  $X'_B(t')$ of the ends of the rod, follow directly from the Lorentz transformation equations 
\begin{eqnarray}
x'&=&\gamma[x(t)+vt]\label{LTx}\\
t'&=&\gamma[t-vx(t)].
\label{LT}
\end{eqnarray}
These can be treated as parametric equations for $x'$ and $t'$ in terms of the rest frame time $t$, 
with $x(t)$ given by Eqs.~(1)-(3) for each of the trajectories.
\\
{\bf Fig.~2a}\\
Particle A has to start further from the rod in figure 2b, so that it will strike the moving rod on its left side  at the same time (in system S$'$) that the right end of the rod  strikes particle B.  This means that, in frame S, 
particle B will strike the right end of the rod before particle A reaches the left end of the rod.
Both ends of the rod will follow Eq.~(\ref{eq:rod}), and particles A and B will follow Eq.~(\ref{xLR}) until 
\begin{equation}
t_B=L/2v,
\end{equation}
at which time particle B strikes the right end of the rod.
Then, particle B will remain at rest temporarily with
\begin{equation}
x_B=L/2,\quad t_B<t<t_B+2v\gamma L,
\end{equation}
having given all of its momentum to the rod.
The time $2v\gamma L$ that particle B remains at rest is determined by the time
$v\gamma L$ that it takes the right end of the rod to come to a momentary stop when the left end of the rod is struck by particle A, and another length of time
$v\gamma L$
when the right end of the rod returns to collide with particle B a second time.  These times are given by Eq.~(20) of Ref.~\cite{jfr}.
After this second collision, the rod remains at rest at its original position $X_R=L/2$, and particle B resumes its motion with velocity $v$.

When particle B hits the right end of the rod at 
$t=t_B$, the left end of the rod starts to follow the curved trajectory\cite{t} given by  Eq.~(19) of Ref.~\cite{jfr} for impulsive acceleration of the front-end of a rigid body:
\begin{equation}
X_L=L/2-\sqrt{L^2+(t-t_B)^2},\quad t_B<t<t_A,
\label{eq:XL}
\end{equation}
where, $t_A$ is given by Ref.~\cite{jfr} to be
\begin{equation}
t_A=t_B+v\gamma L.
\end{equation}
Because the left end of the rod is momentarily stopped by particle A at time $t_A$, its right end follows the trajectory
\begin{equation}
X_R=\sqrt{L^2+(t-t_A)^2}+L/2-\gamma L,\quad t_B<t<t_A.
\label{eq:XR}
\end{equation}

Particle A originally moves  with velocity v on a path that strikes the left end of the rod at time $t_A $:
\begin{equation}
x_A=-\gamma L(1+v^2)+vt,\quad t<t_A.
\end{equation}
Although $t_A$ is later than $t_B$ in frame S, the times $t_A'$ and $t_B'$ are equal in frame S$'$, as can be seen by Lorentz transforming the times from frame S to frame S$'$.  We see that, although particles A and B strike the rod simultaneously in frame S$'$ where the rod is moving, these equal and opposite impulses do not cancel because they are not simultaneous in the rest system of the rod.

When particle A strikes the rod in frame S, particle A and the rod rebound with the same momentum they had before the collision. This means that particle A will rebound with the straight line trajectory
\begin{equation}
x_A=L(1-1/\gamma)-vt,\quad t>t_A,
\end{equation}
and, the left end of the rod will follow the trajectory
\begin{equation}
X_L=-2L+\sqrt{L^2+(t-t_R)^2},\quad t_B<t<t_B+2\gamma v.
\label{eq:XL2}
\end{equation}
This is just the reverse path to that given by Eq.~(\ref{eq:XL}), and ends with the left end of the rod coming to final rest at its original position, $x = - L/2 $.

The final result is that the rod is back at rest and particles A and B have reversed their velocities, just as in Fig. 1a. However there has been considerable zig-zag movement, because the original impacts of particles A and B were not simultaneous in the rod's rest system.

{\bf Fig.~2b:}\\ 
In system S$'$, the right end of the rod and particle B follow the same initial trajectories as in Fig. 1b
until the time  
\begin{equation}
t'=t'_B=(\gamma L/2)(1+v^2).
\end {equation}
At that time the right end of the rod strikes particle B, and comes to a momentary halt.
The left end of the rod initially moves at constant velocity with
\begin{equation}
X'_L=-L/(2\gamma)+vt',\quad t'<t'_B-\gamma v L.
\end {equation}
Then, as described in Ref.~[2], the left end of the rod starts to slow down, and follows the curved trajectory
\begin{equation}
X'_L=L/2\gamma-L+\sqrt{L^2+(t-t'_B)^2},\quad t'_B-\gamma v L<t'<t'_B.
\label{eq:XpL2}
\end{equation}
This results in the rod being momentarily at rest at time $t'_B$, with its rest length $L$.

Particle A has been moved to the left of the starting position it had in Fig. 1b, and 
follows the initial trajectory
\begin{equation}
x'_A=-L+\frac{2v}{1+v^2}t',\quad t'<t'_A=t'_B.
\label{eq:XpL3}
\end{equation}
Thus, in frame S$'$, particle A strikes the left end of the rod at a time, $t'_A$, which is the same as the time, $t'_B$, that the right end of the rod strikes particle B.  The first left and right impacts on the rod are simultaneous in its moving frame S$'$, but not in it rest frame S.
Particle A is brought to rest by its impact on the rod, and remains at the position
\begin{equation}
x'_A=\gamma L-L,\quad t'>t'_A.
\end{equation}

After being struck by particle A at time $t'_A$, the ends of the rod follow trajectories that 
follow directly from the Lorentz transformation equations of Eqs.~(\ref{LTx}) and (\ref{LT})
treated as parametric equations for $x'$ and $t'$ in terms of the rest frame time $t$, 
with $x(t)$ given by Eqs.~(\ref{eq:XL}) and (\ref{eq:XR}) for each of the trajectories.
These trajectories end at $t'=t'_B+2\gamma v L$, after which time, each end of the rod follows its original trajectory.

After being struck by the right end of the rod, particle B moves with constant velocity until it is struck a second time by the rod:
\begin{equation}
x'_B=L/2\gamma+vt',\quad t'_B<t'<t'_B+2\gamma v L.
\end{equation}
After this collision, particle B moves with the initial velocity of particle A: 
\begin{equation}
x'_B=L/2\gamma+\frac{2vt'}{1+v^2},\quad t'>t'_B+2\gamma v L.
\end{equation}


\begin{thebibliography}{9}
\bibitem{t}When we refer to 'curved' or 'straight' for a trajectory, we mean the path followed in the space-time $x,t$ plane. All the trajectories are straight lines in physical space.
\end{thebibliography}
\end{document}